\documentstyle[twocolumn,prl,graphicx,aps]{revtex}
\begin{document}
\draft
%%%%%%%%%%%%%%%%%%%%%%%%%%%%%%%%title%%%%%%%%%%%%%%%%%%%%%%%%%%%%%%%%%%%%%%%%%%
\title{
The phase transition between dimerized-antiferromagnetic and 
uniform-antiferromagnetic phases in impurity-doped spin-Peierls 
cuprate CuGeO${}_3$
}
\author{
T. Masuda,\footnote{E-mail: tt66497@hongo.ecc.u-tokyo.ac.jp} 
A. Fujioka, Y. Uchiyama, I. Tsukada, and K. Uchinokura
}
\address{
Department of Applied Physics, The University of Tokyo, 
7-3-1 Hongo, Bunkyo-ku, Tokyo 113-8656, JAPAN 
}
\date{January 9, 1998}

\maketitle

%%%%%%%%%%%%%%%%%%%%%%%%%%%%%%%abstract%%%%%%%%%%%%%%%%%%%%%%%%%%%%%%%%%%%%%%%%
\narrowtext
\begin{abstract}
We discovered a first-order phase transition between 
dimer-\break zed-antiferromagnetic and uniform-antiferromagnetic phases 
in impurity-doped spin-Peierls (SP) cuprate 
Cu${}_{1-x}$Mg${}_x$GeO${}_3$. 
As impurity concentration increases, linear reduction of 
SP transition 
temperature ($T_{SP}$) and linear increase of N\'eel 
temperature ($T_N$) are observed up to $x \simeq$ 0.023. 
At this critical concentration ($x_c$) SP transition suddenly 
disappears and $T_N$ jumps discontinuously. 
The peak of the susceptibility at $x_c$ around $T_N$ is not so 
sharp as those at other concentrations, which indicates the phase 
separation of low and high concentration phases. 
These results indicate the existence of the first-order phase 
transition between dimerized-antiferromagnetic and 
uniform-antiferromagnetic long-range orders. 
\end{abstract}
\narrowtext

%%%%%%%%%%%%%%%%%%%%%%%%%%%%%%introduction%%%%%%%%%%%%%%%%%%%%%%%%%%%%%%%%%%%%%

%\section{Introdction}

Since Hase, Terasaki, and Uchinokura discovered the first 
inorganic spin-Peierls (SP) material CuGeO${}_3$ in 
1993~\cite{hase1}, this material has attracted much attention. 
Soon after that substitution effect of nonmagnetic impurity 
(Zn${}^{2+}$) for Cu${}^{2+}$ was studied by Hase 
{\it et al.}~\cite{hase2} and a new magnetic phase was discovered 
below the spin-Peierls transition temperature ($T_{SP}$), which 
turned out to have antiferromagnetic long-range order 
(AF-LRO)~\cite{oseroff,hase3}. 
The neutron scattering experiments were studied on 
Si-~\cite{regnault} and Zn-doped~\cite{sasago1,martin} CuGeO${}_3$ 
and both dimerization superlattice peak and AF magnetic peak were 
observed. 
Fukuyama {\it et al.} explained the coexistence of the 
dimerization and the AF-LRO in CuGe${}_{1-x}$Si${}_x$O${}_3$ using 
phase Hamiltonian~\cite{fukuyama}. 
According to their theory, both dimerization and 
$\langle S^{z}\rangle$ of spins on Cu${}^{2+}$ ions have 
spatially inhomogeneous distribution. 
Recent $\mu$SR study on Zn- and Si-doped CuGeO${}_3$ indicated the 
spatial inhomogeneity of $\langle S^{z}\rangle$ of spins on 
Cu${}^{2+}$ ions in AF-LRO phase~\cite{kojima}, which supports 
Fukuyama {\it et al.}'s theory. 

Transition temperature vs impurity concentration  ($T$-$x$) phase 
diagrams have been proposed on Zn- and Si-doped 
CuGeO${}_3$~\cite{sasago1,coad,renard}. 
In both cases N\'eel temperature ($T_N$) increases gradually, 
reaches its maximum and decreases moderately. The $T_{SP}$ 
decreases linearly as $x$ increases. 
However, in the case of Zn-doped CuGeO${}_3$ $T_{SP}$ was reported 
to have a plateau in highly doped region~\cite{sasago1}, while in 
the case of Si-doped CuGeO${}_3$ the corresponding plateau was not 
observed~\cite{renard}. 
The $T$-$x$ phase diagram is controversial in the relatively 
highly doped region and the study on the substitution by other 
species of impurities is needed. 

In this paper we study $T$-$x$ phase diagram in 
Cu${}_{1-x}$Mg${}_x$GeO${}_3$  in detail and report (a) the clear 
disappearance of $T_{SP}$, the 
corresponding jump of $T_N$ and (b) the 
existence of different AF-LRO's with and 
without the lattice dimerization. 

%%%%%%%%%%%%%%%%%%%%%%%%%%%%%%experimental details%%%%%%%%%%%%%%%%%%%%%%%%%%%%%

%\section{Experimental details}

All single crystals were grown by a floating-zone method. 
A typical dimension of the grown crystals is about 4-5 mm in 
diameter and about 4-8 cm in length. 
The true concentration of impurity $x$ was determined by 
Inductively Coupled Plasma Atomic Emission Spectroscopy (ICP-AES). 
We use Ar gas as plasma source and perform quantitative 
analysis by calibration curve method. 
The $x$ for Mg is over 80\% of nominal concentration $x_{nom}$ for 
$0 \le x_{nom} < 0.1$. This is in contrast to that in  Zn-doped 
CuGeO${}_3$~\cite{martin}, where the ratio is below 80\%. 
This means that Mg is more easily doped to Cu-site and is expected 
to be a 
more adequate impurity than Zn for the study of the substitution 
effect of nonmagnetic ions. 
This is one of the reasons why we have reinvestigated the $T$- $x$ 
phase 
diagram in detail in Mg-doped CuGeO${}_3$. 
The absence of impurity phase or structure change with $x$ was 
confirmed 
by x-ray diffraction after pulverization of the single crystals 
at room temperature. 

Measurements of DC magnetic susceptibility were performed with 
commercial 
SQUID  magnetometer ($\chi$-MAG, Conductus  Co., Ltd.) for 34 
samples 
($0\le x<0.089$).

%%%%%%%%%%%%%%%%%%%%%%%%%%%%%%experimental results%%%%%%%%%%%%%%%%%%%%%%%%%%%%%

%\section{Experimental Results and Discussion}

\begin{figure}
\includegraphics*[width=8cm]{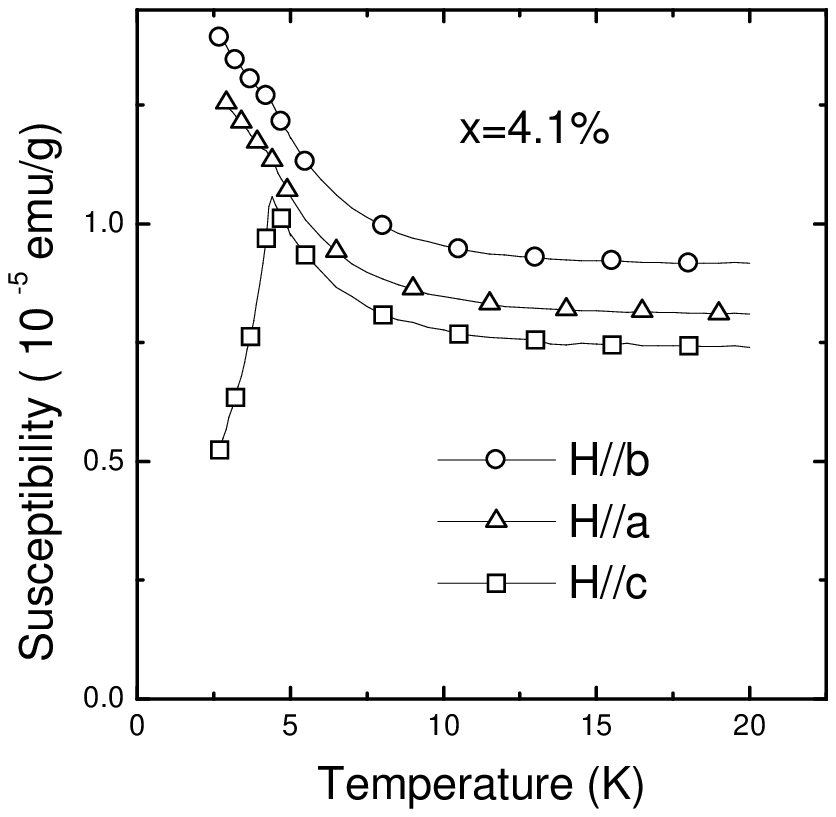}
\vspace{5mm}
\caption{
Temperature dependence of susceptibility on the 
sample of 
$x = 0.041$ in the field applied parallel to the three principal 
axes. 
N\'eel transition at 4.4 K is observed. 
}
\label{fig1}
\end{figure}

The susceptibility changes anisotropically at low temperatures as 
shown 
in Fig.~\ref{fig1}. 
We can see that (a) Mg can be doped~\cite{ajiro,weiden}, 
(b) Mg-doping induces AF-LRO as in Zn-~\cite{hase3}, 
Ni-~\cite{coad,koide2}, Mn-~\cite{oseroff}, and Co-doped 
CuGeO${}_3$~\cite{anderson}, and (c) the magnetic easy axis is 
along the $c$ axis below $T_{N}$, which is the same as in the case 
of Zn-doped CuGeO${}_3$~\cite{hase3}. 
Both $T_N$ and $T_{SP}$ were determined from the crossing points 
of linear functions fitted to the susceptibility in applied field 
parallel to the $c$ axis ($\chi _{c}(T)$) above and below the 
transitions.

Fisher reported that the magnetic heat capacity 
of a `simple' antiferromagnet 
is proportional to $\partial (\chi_{\parallel} T)/\partial T$ 
and $T_N$ is best determined by the 
maximum in $\partial (\chi_{\parallel} T)/\partial T$
($\chi_{\parallel}$ is the susceptibility 
along the easy axis, which corresponds to $\chi_c$ 
in this case)~\cite{fisher}. 
The maximum in $\chi_{\parallel}$, therefore, occurs at a temperature 
slightly higher than $T_N$. 
We analyzed some of the data by this way 
and get, {\it e.g.} $T_{N}$ = 4.3 K for the sample of $x = 0.035$, 
(Note that the temperature step was 0.1 K). 
This value is closer to the $T_N$, 4.2 K, which was determined from 
our heat capacity measurement than the value, 4.5 K, determined 
from the maximum in $\chi_c$. 
In the present paper, however, we determine $T_N$ by the method 
described above 
because Fisher's method can be applied to a `simple' antiferromagnet, 
to which the low-concentration antiferromagnetic phase in 
Mg-doped CuGeO${}_3$ does not belong 
and because the change of $T_N$ with $x$ but not 
the absolute value of $T_N$ 
is essential in the present study. 

\begin{figure}
\includegraphics*[width=8cm]{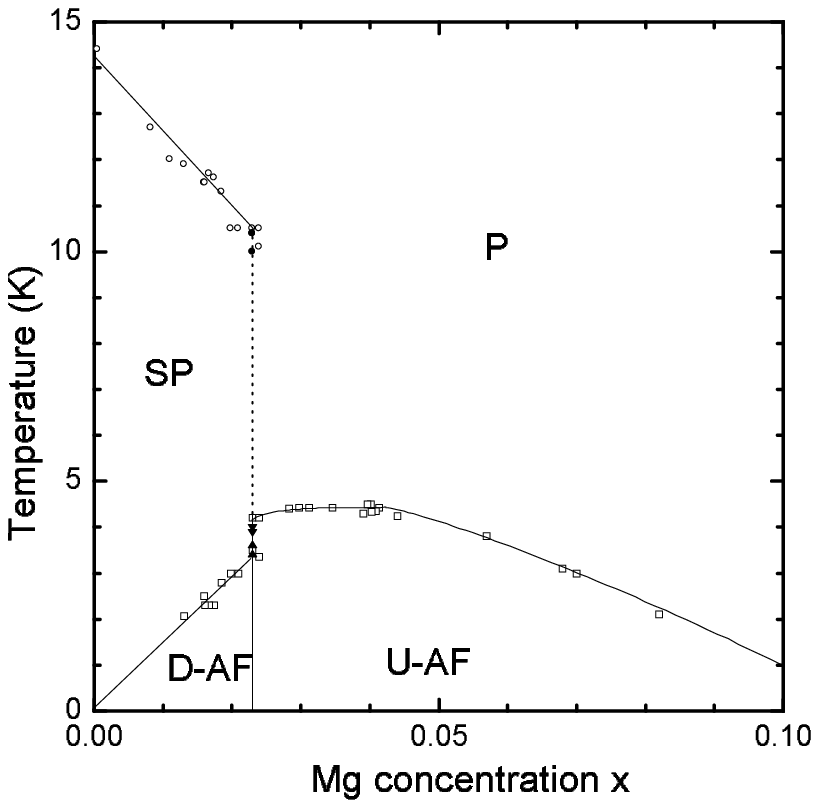}
\vspace{5mm}
\caption{
The $T$-$x$ phase diagram of Cu${}_{1-x}$Mg${}_x$GeO${}_3$. 
Circles and squares indicate $T_{SP}$ and $T_N$, respectively. 
At $x = 0.023$ jump of $T_N$ and sudden disappearance of 
$T_{SP}$ are observed. 
Filled triangles represent $T_1$ (upward triangle) and $T_2$ 
(downward one) 
at $x_c$, which are determined 
as shown in Fig.~\protect\ref{fig3}(c). 
SP and P mean spin-Peierls and paramagnetic states. 
The meaning of D-AF and U-AF are explained in the text. 
}
\label{fig2}
\end{figure}

Figure \ref{fig2} shows Mg concentration dependence of $T_{SP}$ 
and $T_{N}$: $T$-$x$ phase diagram. 
$T_{N}$ increases from 3.4 K to 4.2 K abruptly at $x \simeq 0.023$ and 
reaches its maximum. 
We define this critical concentration as $x_c$. 
$T_N$ has a plateau at $x_c < x {<\atop\sim} 0.04$ and decreases 
smoothly at $x {>\atop\sim} 0.04$. 
The N\'eel transition was not observed in the sample of $x = 0.089$ above
1.9 K. 
On the other hand $T_{SP}$ reduces linearly from 14.2 K of pure 
CuGeO${}_3$ and suddenly 
disappears at $x_c$ around 10 K and is not observed at $x > x_c$. 

\begin{figure}
\includegraphics*[width=8cm]{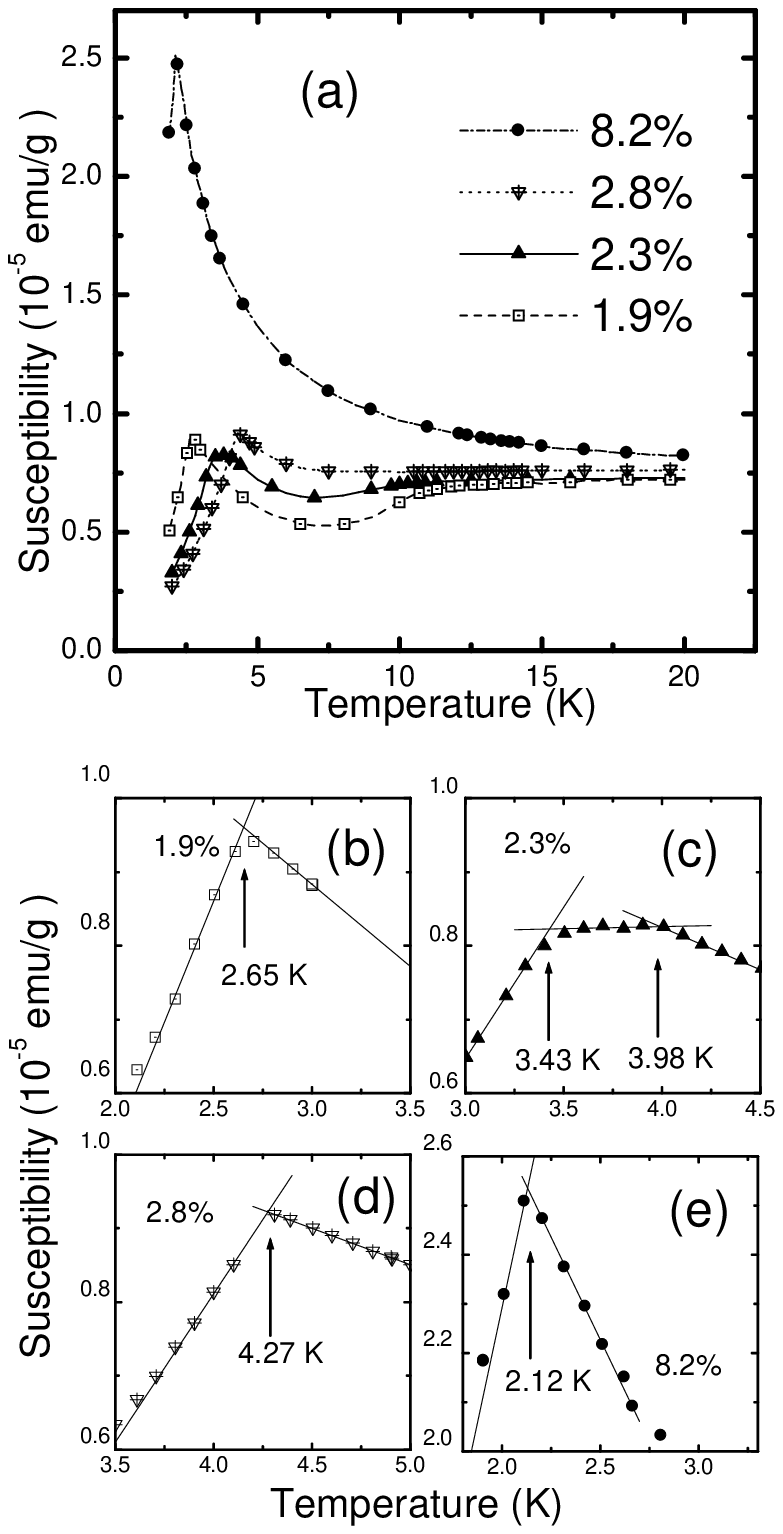}
\vspace{5mm}
\caption{
(a) $\chi_c (T)$ of Cu${}_{1-x}$Mg${}_x$GeO${}_3$ 
with $x$ = 0.019, 0.023 ($\simeq x_c$), 0.028, and 0.082. 
(b) $\sim$ (e) $\chi_c (T)$ near $T_N$. 
While below and above $x_c$ the peaks are sharp as shown in (b), 
(d), and (e), 
at $x \simeq x_c$ the peak is broad as shown in (c). 
We determined the transition temperatures $T_1$ and $T_2$ at 
$x = x_c$ as crossing points of 
fitted three linear functions of $T$.  $T_1$ = 3.4 K and 
$T_2$ = 4.0 K in (c). 
}
\label{fig3}
\end{figure}

Figure~\ref{fig3}(a) shows $\chi _c(T)$ of Mg-doped CuGeO${}_3$ 
($x = 0.019,$ $0.023(\simeq x_c),$ $0.028$ and $0.082$). 
Figures~\ref{fig3}(b)$\sim$(e) show the same data 
as in Fig.~\ref{fig3}(a) near $T_N$. 
Below and even above $x_c$ 
sharp transitions are 
observed in Figs.~\ref{fig3}(b), (d), and (e). 
The measurements were done in the steps of 0.1 K and the 
broadening of the 
peaks was not observed. 
Therefore the errors of $T_N$ is less than 0.05 K at these $x$'s. 
At $x_c$, however, the broadening of the peaks is observed as shown 
in Fig.~\ref{fig3}(c). 
This behavior indicates the existence of two transition 
temperatures $T_1$ and $T_2$, which is caused by a phase 
separation into a low and high concentration phases. 
It is noted that a phase separation always appears in the case of 
a first-order phase transition. 
Here we analyzed the data by fitting three linear functions of $T$ 
and 
determined crossing points as $T_1$ and $T_2$. 
These are 3.43 K and 3.98 K at $x = 0.023$ (Fig.~\ref{fig3}(c)). 

According to the susceptibility data
we can explain Fig.~\ref{fig2}\ as follows. 
First the jump of $T_N$ at $x = x_c$ indicates that 
AF-LRO at $x < x_c$ and $x > x_c$ belong to essentially different 
phases and there is a distinct phase transition between them. 

Second the disappearance of the SP transition at $x_c$ implies 
that the 
lattice dimerization is absent, {\it i.e.}, the lattice is uniform 
in the region of $x > x_c$. 
Therefore it is inferred that at $T < T_N$ the lattice remains 
uniform. 
We define this phase as the uniform-antiferromagnetic 
phase (U-AF phase). 
The U-AF phase is supposed to be classical; there is no 
spatial inhomogeneity of $\langle S^z\rangle$ 
of the spins on Cu${}^{2+}$ ions. 
In the sample of $x = 0.041$ the absence of dimerization 
was confirmed by neutron diffraction measurement 
down to 1.3 K~\cite{nakao}. 
On the other hand in the region of 
$x < x_c$ the lattice is dimerized below $T_{SP}$. 
It is expected that the lattice is dimerized below $T_N$, 
which was also confirmed by neutron scattering 
measurement on the sample of $x = 0.017$~\cite{nakao}. 
There should be spatial inhomogeneity of Cu 
spins as is claimed so far in 
Si-doped CuGeO${}_3$~\cite{fukuyama}. 
Here we define this phase as dimerized-antiferromagnetic phase (D-AF phase). 

Last the broad peak of $\chi_c (T)$ in the 
sample of $x \simeq x_c$ indicates the transition 
from D-AF to U-AF phases is the first order as $x$ is varied. 
The displacement of Cu${}^{2+}$ ion, $\delta$, 
from a uniform lattice changes abruptly from 
finite value to zero at $x = x_c$. 

As briefly mentioned previously the absence of the structure change with $x$
was confirmed by x-ray diffraction at room temperature.

Once we know the presence of the first-order phase transition 
in Cu${}_{1-x}$Mg${}_x$GeO${}_3$, it becomes important to review $T$-$x$
phase diagrams 
of Cu${}_{1-x}M_x$GeO${}_3$ ($M$ = impurity). 
In the case of Zn-doped CuGeO${}_3$, the absence of $T_N$ between 
3.0 and 4.2 K at $x \sim 0.017$ was observed (see Fig.~\ref{fig2} 
of Ref.~\cite{koide1}). 
This suggests that the first-order phase transition also exists in 
this system.
However the jump of $T_N$ and the corresponding vanishment of the 
SP transition have not been clearly confirmed so far.
This, we think, is because the distribution of Zn in the sample is 
not 
so uniform as that of Mg and the phase boundary was disturbed by 
this effect. 
In the case of Ni-doped CuGeO${}_3$, sudden disappearance of 
$T_{SP}$ and abrupt increase of $T_{N}$ from 2.5 K to 3.6 K 
at $x = 0.020$ were clearly observed~\cite{koide2}. 
This corresponds to the phase transition observed in
Cu${}_{1-x}$Mg${}_x$GeO${}_3$.
However, the behavior is more complex owing to the difference of 
the easy axis (nearly parallel to the $a$ axis in Ni-doped 
CuGeO${}_3$)~\cite{coad,koide2}\ and the detail will be discussed
separately~\cite{koide2}. 

The plateau of $T_{SP}$ at relatively large $x$ is observed 
by the neutron diffraction, but only very weakly by the 
susceptibility measurement, in the case of Zn-doped 
CuGeO${}_3$~\cite{sasago1,martin}. 
This may also be explained by spatial variation of Zn 
concentration. 
Scattering from low concentration ({\it i.e.}, dimerized) 
region can be observed by the neutron diffraction even 
though the volume of that region is small. 
On the other hand the susceptibility measurement detects 
the average property of a sample.

From the above discussion the first-order phase transition seems 
to be universal for all dopants at least in the case of doping to 
Cu site. 

The results of antiferromagnetic 
resonance~\cite{hase4} and of angular 
dependence of magnetization vs magnetic field~\cite{koide2} on 
Zn-doped 
(4\%) CuGeO${}_3$ were explained very well using mean-field 
sublattice 
model~\cite{hase4,koide2}. 
This may also be explained by the fact that the magnetic phase of 
these 
samples at $x > x_c$ is perfectly classical U-AF. 
Different behaviors are expected in the samples at $x < x_c$. 

While D-AF phase has AF-LRO characteristic to SP state, 
U-AF has classical AF-LRO, which arises because the interchain 
exchange 
interaction of CuGeO${}_3$ is not so weak~\cite{nishi} as 
that of other typical organic SP materials~\cite{bray,huizinga}. 
In other words, if SP transition had not occurred in CuGeO${}_3$, 
even pure CuGeO${}_3$ would be a classical AF material. 
The disappearance of lattice dimerization may induce the phase 
transition 
from D-AF to U-AF phases through spin-lattice coupling. 
The energies of D-AF and U-AF phases including both spin and 
lattice 
should be calculated in the ground state and sudden disappearance 
of 
SP transition should be also explained. 

M. Weiden {\it et} {\it al.} also reported 
$T$-$x$ phase diagram of Mg-doped CuGeO${}_3$ 
from susceptibility measurements~\cite{weiden}. 
But they have neither shown the susceptibility data nor 
clarified how they determined the concentration $x$. 
On the other hand we, first, checked that of 
emission spectra 
of Cu (= 327.396 nm), Ge (=209.423 nm), and Mg (= 279.553 nm) 
do not interfere with each other in 
ICP-AES measurement. 
Second we made sure that the detection limit of 
the intensity of Mg spectrum 
is much smaller than the intensity of our usual 
samples (about 4 mg of Cu${}_{1-x}$Mg${}_x$GeO${}_3$ 
for $0.001 {<\atop\sim} x$) for ICP-AES measurement. 
Third we performed quantitative analysis on 
a few nearest neighboring samples and 
we confirmed that the fluctuation of $x$ is within 
0.001. 
The detailed composition analysis and the good choice 
of impurity make the discovery of the present phase transition 
possible. 

As future problems the properties of the two phases should be 
studied 
close to the first-order phase boundary using various kinds 
of physical measurements: 
neutron diffraction and neutron inelastic scattering, $\mu$SR, 
specific heat, 
and x-ray diffraction at low temperatures. 
As to Zn-doped CuGeO${}_3$ we are planning to reinvestigate 
the phase diagram around $x \sim 0.017$ in detail and to 
clarify whether the jump of $T_N$ really exists or not. 
Further theoretical explanation of the phase transition is also 
needed. 
Another problem is whether the phase transition exists in Si-doped 
CuGeO${}_3$, in other words, whether it is unique to the  doping 
to Cu-site or not. 
Detailed studies on the $T$-$x$ phase diagram of 
CuGe${}_{1-x}M_x$O${}_3$ 
are needed. 

%%%%%%%%%%%%%%%%%%%%%%%%%%%%%%%%%summary%%%%%%%%%%%%%%%%%%%%%%%%%%%%%%%%

%\section{Summary}

In summary we studied in detail the $T$-$x$ phase diagram of 
Cu${}_{1-x}$Mg$_x$GeO${}_3$ 
and discovered a first-order phase transition between D-AF and 
U-AF phases. 
At $x_c$ $\delta$ changes from finite value to zero and spatial 
distribution of $\langle S^z\rangle$ also changes from 
inhomogeneous to uniform distributions. 
The transition seems to be universal for the doping to Cu site and 
we can explain some of the unsolved problems in impurity-doped 
CuGeO${}_3$ 
by this $T$-$x$ phase diagram. 

We thank G. Shirane, Y. Fujii, H. Nakao, M. Nishi, and K. Hirota 
for neutron diffraction measurement on Mg-doped CuGeO${}_3$. 
This work was supported in part by Grant-in-Aid for Scientific 
Research (A),
Grant-in-Aid for Scientific Research on Priority Area 
``Anomalous Metallic State near the Mott Transition'', 
Grant-in-Aid for COE Research 
``Phase Control of Spin-Charge-Photon (SCP) Coupled System''
from the Ministry of Education, Science, Sports, and Culture,
and NEDO (New Energy and Industrial Technology Development 
Organization) International Joint Research Grant.

%%%%%%%%%%%%%%%%%%%%%%%%%%%%%%%%reference%%%%%%%%%%%%%%%%%%%%%%%%%%%%%%%%%%%%

%%%%%%%%%%%%%%%%%%%%%%%%%%%%%%%%%figure%%%%%%%%%%%%%%%%%%%%%%%%%%%%%%%%%%%%%%

\end{document}